\documentstyle[11pt,newpasp,twoside,epsf]{article}
\def\edcomment#1{\iffalse\marginpar{\raggedright\sl#1\/}\else\relax\fi}
\reversemarginpar

\begin{document}
\title{CNO abundances in the Damped Ly$\alpha$ clouds}
 \author{Paolo Molaro}
\affil{Osservatorio Astronomico di Trieste-INAF, Via G.B. Tiepolo 11, }

\begin{abstract}
Damped Ly$\alpha$  clouds   provide information on the chemistry  of distant and metal poor regions of the universe.
In these clouds for   observational difficulties N and O are among the last elements   to be  measured, and C 
 has  still to  be  measured with precision. In combining  the extant CNO abundances,
  we  avail of a sample which includes   33 N  measurements or significant limits,  a dozen of   O measurements, and only few  tentative measurements for C.
O is found  to track both S and Si rather  closely   without signatures of any peculiar behaviour and the few  [O/Zn] ratios available show   
a mild $\alpha$/iron-peak enhancement    which reinforces earlier evidence. The few  tentative
C measurements point to  roughly solar  [C/Zn] and [C/Si] ratios.

  N abundances are rather complex with  N/$\alpha$  ratios showing  a bimodal distribution.    
The majority of the values form a first plateau    at   [N/Si]=$-$0.82 ($\pm$ 0.13) and  about 25 \%   form a second plateau  at 
[N/Si]= $-$1.45 ($\pm$ 0.05).
    The ${\it high}$  values are at the level  of the Blue Compact Dwarfs but they extend further 
  towards lower  metallicities.
The {\it lower} plateau  is  a   new feature of  DLAs with no   other counterparts.
The two plateaux are found for different values of the N abundance with the transition   at [N/H]$\approx$ -3.0.
We argue  that this   results  from the delayed release of N between massive  and intermediate mass stars
with the lower plateau produced by massive stars (M$\ge$ 8 M$_{\sun}$ only, which  accounts for the small dispersion.  
The  systems on the lower plateau  are  relatively  young
and   imply 
  a    continuous formation of the DLAs. 
  While the interpretation of the N/O in the DLAs needs to be confirmed,  these observations could ultimately prove 
crucial for the interpretation of the early  N nucleosynthesis.

\end{abstract}

\section{Introduction}

 In the previous ESO workshop  dedicated to the {\it Production and Distribution of C,N,O elements} held in Garching in May 1985   
    the information on CNO  elements  coming from absorption systems along distant QSO  were 
  just at the starting point  and the first   tentative results   were   reviewed by Max Pettini.
We had to wait for one decade for the first N measurement in a damped Ly$\alpha$ system (DLA), even longer  
for  O, and  the   time for C has  probably not  arrived yet.

DLAs are intervening absorption systems  with large hydrogen column densities 
 particularly appropriate for the determination of accurate chemical abundances.
 Typical abundances  are  $-$2.5$<$ [Fe/H]$<$ $-$1   
   with  a moderate chemical evolution. The increase in metallicity 
 is of the order of 0.3 dex   per unit of decreasing redshift  and suggests that DLAs  do not trace 
the  population  of galaxies responsible for the bulk of star formation in the universe.
 A  floor of abundances at about -2.5  
  is observed at all redshifts      up 
to   z=5.28 (Songaila et al 2002) quite  suggestive of a prompt enrichment at very early phases. 
 At redshif  5 the age of the universe  is  about one billion years and there is   hope to find  young systems where we can better 
 recognize  the imprints of the first generation of stars.

 The observed abundances are gas phase abundances and  in general  suffer from  possible dust and ionization corrections. 
  The 
 large neutral hydrogen column densities of the DLAs shield  the clouds from the IGM field effectively. New  evidence comes 
 from the recently measured Ar   which is particularly sensitive to ionization effects.
  The [Ar/Si] is often very close to solar
 or mildly deficient, which rules out significant ionization (Molaro et al 2001, Vladilo et al 2003).
  Dust is definitely present in the DLAs as 
  revealed by a number of indications such as
 the systematic  deficiency of Fe compared with Zn, the anti-correlation between refractory elements 
 and  the elemental column density,  
 the correlations between [Fe/Zn] and H$_2$   and  the correlation between [Si/Fe] and [Zn/Fe] (Molaro 2001 and references therein).
  CNO elements are all mildly  refractory  and are virtually free from this complication (Savage \& Sembach 1996).

  The determination of the   abundances of  the CNO elements have   proven to be very challenging. This   because the most accessible 
lines for CNO are either saturated, despite the low metallicities of the DLAs, 
 or they fall in spectral regions strongly contaminated by the neutral H absorptions  of the  Ly$\alpha$ forest.
The OI 1302 \AA~  line is generally nicely placed redwards the Ly $\alpha$ forest but 
 is  always found strongly saturated.   Conversely  the  OI 1355 \AA~  has never been detected 
 and provides, together with the previous line,  high and low  bounds to the oxygen abundance.
  A similar situation holds for C where the only suitable lines are the CII resonance 
 lines at 1334.53 \AA~   and CII 1036.33 \AA~  
 which are heavily saturated in all   DLAs.
 A more favourable circumstance is found for NI which has several multiplets offering a   relevant dynamical range 
 in the line strenghts. However, the more accessible  NI multiplets are  
     $\lambda\lambda$    1134.16, 1134.42,  1134.98 \AA~  and  1199.55, 1200.22, 1200.71 \AA~    which  
 fall in  the Ly$\alpha$ forest.
 
 \section{Oxygen}
    
First measurements of O were performed  at last through observations of    OI 925 \AA~  and OI 950 \AA~  
     in the Ly-$\alpha$ forest,     associated with the z$_{abs}$=3.39 DLA towards QSO 0000-2621 
 (Molaro et al 2000).
 Partial   contamination by    hydrogen clouds is taken into account using  a model for the 
  absorber constructed on   lines falling redwards the Ly $\alpha$ forest.  
This approach   has been followed   in the DLA  at z$_{abs}$=4.49  towards the  J 0307-4945
 (Dessaugues-Zavadsky et al (2001),  and 
  towards QSO
 0347-381 through the measurement of a dozen    OI lines (Levshakov et al 2002).
    Probably the best   available measure due to the simplicity of the absorber
 and to the quality of Keck data   has been realized by Prochaska et al (2001)
    in the DLA at z=2.844 towards QSO 1946+76 
   were several OI lines have been detected  allowing an OI measure of log(OI)=14.819 $\pm$ 0.007.
 Pettini et al (2002) targeted few systems with low elemental abundances and low HI column density and succeded 
  in measuring O from an almost unsaturated     OI 1302 \AA~   in a couple of  DLAs.
OI  measurements have been provided for  other DLAs towards  QSO 1202-0725 (D'Odorico et al 2003)    QSO 2059-360 (Dessaugues-Zavadsky 
et al 2003),   QSO 2243-6031, QSO 1104-1805     (Lopez et al 1999, Lopez et al 2002).  A total of  a dozen  of systems with some information 
  on the O abundance is presently available.

The [O/Zn] and [O/Fe] are shown in Fig 1 with filled and empty circles respectively,    
and the solar O abundance  taken from the recent 
revision by Holweger (2001).  O and Zn     do not require any dust correction and   two   [O/Zn]
 reveal   a moderate enhancement in $\alpha$-elements at the level of 0.2 dex or so, while in one case the ratio is of 
 $\approx$ 0.6 dex.  There are also few  [O/Fe]$\approx$ 0.2
  which  give   stringent values. Most likely these systems are   dust-free, as   QSO 0000-2621 where [O/Zn]$\approx$[O/Fe].
    If dust is present  the effect is  to  increase the intrinsic Fe abundance and further lower
the [O/Fe] ratio. Overall there is  a low overabundance of O relative to Zn, which is   consistent with  previous indications
obtained from  [S/Zn] measurements  or from [Si/Fe] once expected dust depletion is accounted for (Centuri\'{o}n et al 2000, Vladilo 2001).  

In the   sample  there are two DLAs with O and   S measured. The [O/S] ratio is $\approx$ 0 showing that there 
 is no  particular behaviour of O with respect to the other  $\alpha$ elements. 
 Si is generally available but   it  is a refractory element and may be partially depleted
onto dust grains.  In  all but two DLAs  the [O/Si] is found $\approx$ 0,   which implies that Si 
is only moderately  locked up into dust grains in   DLAs.
    If this is the case it is rather interesting to consider the
behaviour of Si with comparison of the undepleted element Zn. The average value of the 26 DLAs  with both elements measured is 
$<$[Si/Zn]$>$=-0.07 $\pm$0.2,  providing additional evidence of a moderate, if any,  enhancement of $\alpha$-elements.
 Overall the low $\alpha$ over iron-peak element ratio  appears as a prominent characteristic of the DLA. Cases of
genuinely $\alpha$ enhancement  exist such as the DLA towards J0347-3819 but they look   as an   exception rather than the rule
(Levshakov et al 2002).

\begin{figure} 
\plotfiddle{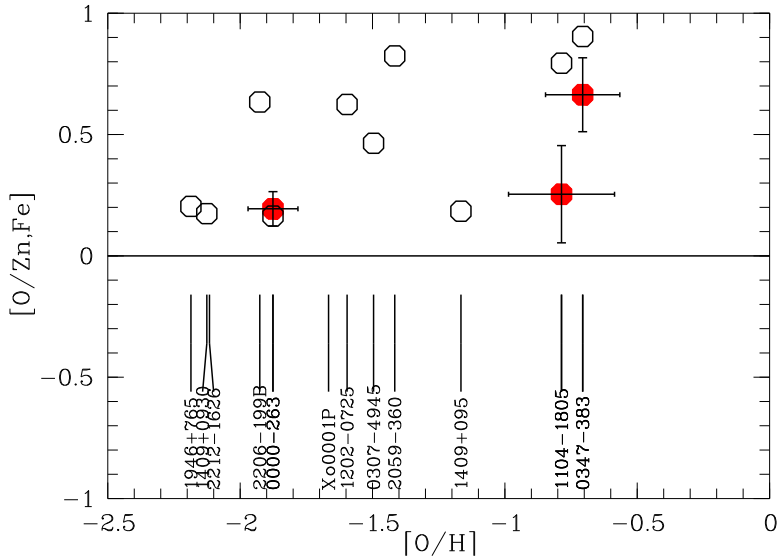}{6cm}{-90}{40}{40}{-200}{230}
\caption{[O/Zn], filled circles, or [O/Fe], empty circles} 
\end{figure}

\section{Carbon}

In  the survey by Prochaska et al 2001 there are  14 lower limits of C abundances, all obtained from the saturated CII 1335 \AA~ line.
 CI is observed in a few cases    but it traces  very cool gas and being  strongly ionised is not very informative of  the total C.
Levshakov et al (2002)  and Lopez et al (2002) through   detailed modelling of all   absorption features in the effort  to  
 account for the saturation of the CII lines,
claimed  measurements with a reasonable error.  
 In these two DLAs  Zn is also measured and for both [C/Zn]$\approx$ 0.25, which suggests  a  rather {\it normal} behaviour.
Highly supersolar [Si/C]  values at the level of 0.5-1.5 dex are  predicted by 
PopIII SNe, hypernovae or pair-instability SNe
of very massive stars (Umeda and Nomoto 2001) for which  there is no  evidence in  DLAs where   [Si/C] is solar.
Better  prospects to   measure C rely on the observations of very high redshift DLAs if   abundances finally  decrease,
or in the class of sub-DLAs (cfr Dessauges-Zavadski, these proceedings). In    the z$_{abs}$=4.383 DLA towards BR 1202-0725 
the CII line is found only moderately saturated and  [C/Fe]=-0.03 ($\pm$0.15)  and [Si/C]=0.15 $\pm$0.11 are
derived  (D'Odorico et al 2003).   In the z$_{abs}$=5.8 DLA
  towards SDSS 1044-0125   identified by Songaila and Cowie (2002), in which  $\log$ N(HI)= 20.5 cm$^{-2}$ and [Fe/H]=$-$2.65,
   the CII 1334 transition  is unsaturated providing a  [C/Fe]=-0.6, which suggests a possible    drop in the C abundance at very high redshift.
  At   redshift about 6 we may be  within  100 million years from  the epoch of star formation,  which puts tight
 constraints to the C build-up processes (Molaro et al 2003).
 
\section{Nitrogen }
 Among the  CNO triplet N was the first   to be observed.    In their first  attempt  Pettini et al (1995)
failed in detecting the element in the DLA towards QSO 2348-146 but established  a very low upper limit 
at   [N/H]$<$$-$3.15, which  
  has been  slightly  improved with Keck observations  to [N/H]$<$$-$3.28     (Prochaska and   Wolfe 1999). 
This  failure was  followed by the    detection of N in the DLA at z$_{abs}$=3.39 towards QSO 0000-2621 and  
 towards QSO  1331+17  with the N abundances     [N/H]=-2.77 $\pm$  0.17,  
    and [N/H]= -2.73,  respectively (Molaro et al 1996,
Kulkarni et al 1996).
 The  three DLAs  showed a difference by $\approx$ 0.6 dex in their [N/Si] ratios and 
     introduced since the beginning the notion of significant scatter in the [N/Si] abundances.
  It is just amazing that this became  evident with the  first two systems and  that   the first one to be 
  observed was a Nitrogen    poor system.

The early   findings were confirmed and extended on larger samples, of which the most important for the number of objects 
are those of
   Centuri\'{o}n et al (1998),  Lu et al (1998),   Pettini et al (2002) and Prochaska et al (2002). In the following we use the 
        complete   and updated compilation of   
   Centuri\'{o}n et al (2002, Table 9) who  contributed with some  new N observations as well.   
The sample of  N observations or limits now comprises 33 systems, out  of which 
  28  have    at least one of    O, Si or S measured and provide  informative ratios.

In  Fig 2   the  [N/$\alpha$]  are plotted versus [$\alpha$/H]. For the $\alpha$ element either of 
 O,   S or  Si are taken in the order. As discussed in the previous section 
 they trace each other rather well so that it does not matter very much which $\alpha$ element is used.
The data for the BCG are also shown in the figure with smaller symbols.

\begin{figure} 
\plotfiddle{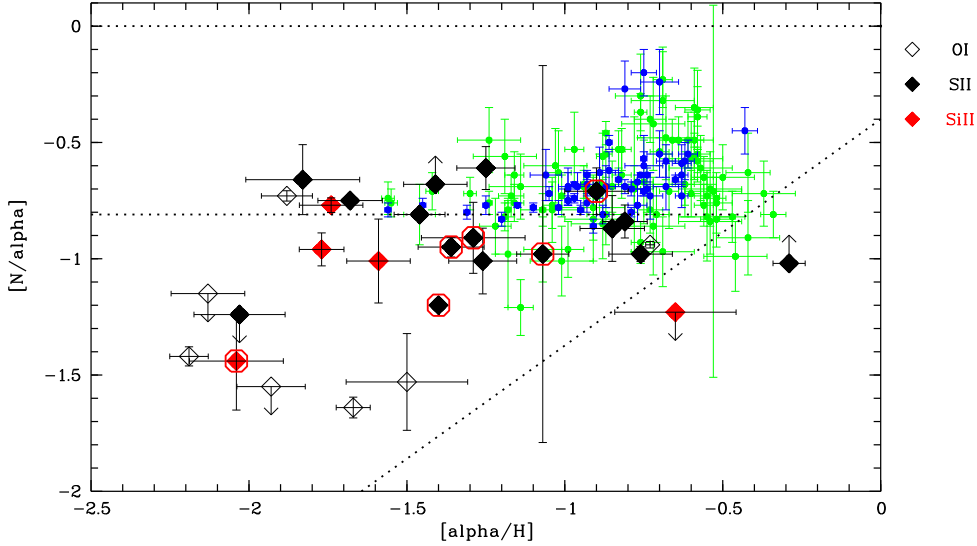}{6cm}{-90}{50}{50}{-200}{280}
\caption{[N/$\alpha$] versus [$\alpha$/H] where O, Si and S are marked by  symbols. The small points are extragalactic HII regions 
 } 
\end{figure}

The  data show a peculiar  pattern. Most of the   sample  form  a 
plateau at [N/Si]$\approx$ $-$0.82 ($\pm$ 0.13), while few points  show   lower N/O ratios.  The 
possibility of a gap   between   low and high values rather than  a pure scatter 
has  been  pointed out by Prochaska et al (2002) who suggested  a bimodal distribution. 
To the two measurements and to one upper limit with low N/O considered originally by Prochaska et al   
we add  the value towards J0307-4945..                                                                                                                                                                               
   Dessauges-Zavadsky (2001), rejected by Prochaska et al (2002) for a possible ionization effect,
    and the  most recent one towards QSO 2059-360 from  Dessauges-Zavadsky et al  (2002).    
Two new     upper limits  quite separate from the high values   and   consistent 
  with a bimodal distribution are
provided by   Pettini et al (  2002).
   The average of the four determinations 
is [N/Si]$\approx$ $-$1.45($\pm$ 0.05), strongly suggestive of  a second plateau even with 
  a narrower dispersion. 
The low N/O values are mostly observed     at the lowest metallicities, but  with an overlap over  the higher  plateau 
 in the  range   -2.0 $<$[O/H]$<$ -1.5.
In   Fig 3  the  plot of   [N/H] versus [Si/H]   clearly  shows   the  primary-like  behaviour of  N in both plateau.

The high N/O values  look as an extension of the   BCG  towards lower metallicities and 
 give definitive  evidence of  a  plateau,  which was not so clear from the  BCG only. 
The sharing of the same N/O between DLAs and BCG  suggest that    the two populations   experienced  similar chemical evolution.
 
 N synthesis is thought to occur in  Intermediate Mass Stars (IMS) (4 to 7 M$_{\sun}$) through 
  Hot Bottom Burning in the  AGB phase. 
The    nucleosynthesis  by IMS  implies a delay  in the N production when compared to that of O 
 as well as  a	sensitivity of N production to the low mass regime  of the  IMF.
 The general interpretation of the observed data is
  that N is primary at low metallicities   and secondary at high metallicity  (Henry et al 
 2000).  The Henry et al   model is able to reproduce the   plateau by reducing the SFR and thus by preventing
  the rapid O build up by massive stars. For high  SFR 
   the elemental build up is fast and the system enters very soon in the secondary 
production regime without undergoing  a primary behaviour at all.
  Thus the presence of an extended  plateau in the N/O suggests  low SFR in the DLA, which 
may also explain the  failure   of imaging   DLA galaxies.
The {\it lower} plateau  is  a   new feature of  DLA with no   other counterparts.
   Izotov et al (1999)    argued that 
 the low N/Si were resulting from an inflated Si column density by  additional  Si formed in the ionized region. 
 However,  the recent direct O measurements provided five  genuine low N/O obtained by means of truly oxygen observations, 
 shown in Fig 2,  which rule out such  possibility.

\begin{figure} 
\plotfiddle{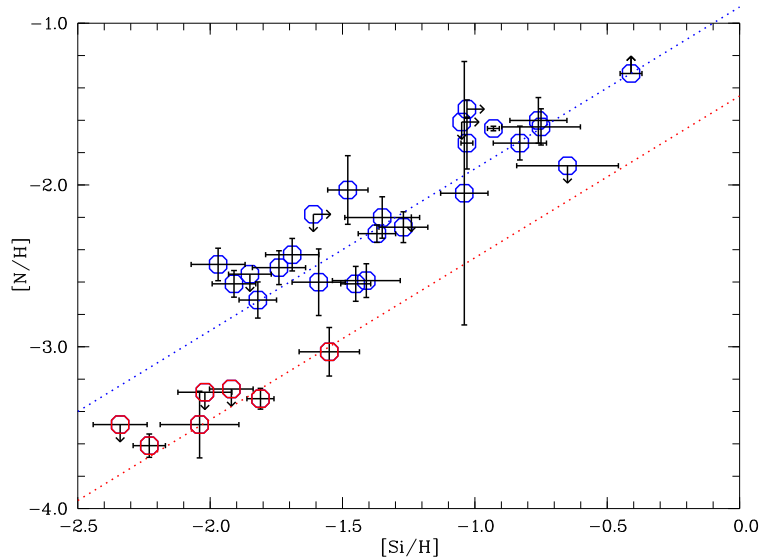}{6cm}{-90}{40}{40}{-200}{230}
\caption{[N/H] versus [Si/H]} 
\end{figure}

\subsection{About the origin of the low N/O  plateau}

Low N/O values have been interpreted in a context of a delayed N released model as the result of a relatively young age 
  ( Molaro 2001, Pettini 2002,) or as being  
  the system observed close to a recent 
episode of star formation  (Lu et al 1998, Centuri\'{o}n et al 1998).
According to   Henry et al  the necessary scale  time to  attain  a  significant N production by  IMS is of
at   least  250 Myrs. It is therefore possible that DLAs 
are catched in the early formation times when the IMS had not yet expelled their N. However,
 as Prochaska et al  pointed out  one should expect to find  values everywhere 
below the plateau  
 and it is difficult to account for the fact that the low N/O  show a characteristic  value at [N/O]=$-$1.4.
  In the low N/O plateau the scatter is so small that  these systems  are   unlikely   in a
transient phase towards  the upper plateau. 
 
 A second possibility to explain the
 low N/O values is a variation in the IMF.  A  top-heavy IMF or  an IMF  truncated below a  mass threshold of $\approx$ 5 M$_{\sun}$
 which exclude  low mass stars      has been suggested by Prochaska et al (2002) .   However, it remains unclear what   produces the  change in the IMF. As discussed by Henry at this meeting 
 it is  rather appealing to think that it is related to the formation of 
   PopIII stars
   
A further possibility   is that we are dealing with   variable N yields as a function of the metallicity since the low N/O values
 occur preferably at the lowest metallicities.     This may   affect the efficiency of the 3rd dredge-up 
 or of the mass-loss rate along the AGB phase 
 somewhat altering the characteristic N yields of the IMS. However, it is not clear how to account
  for the fact that  both  high and low N/O values 
 occur at  equal metallicities.

\begin{figure} 
\plotfiddle{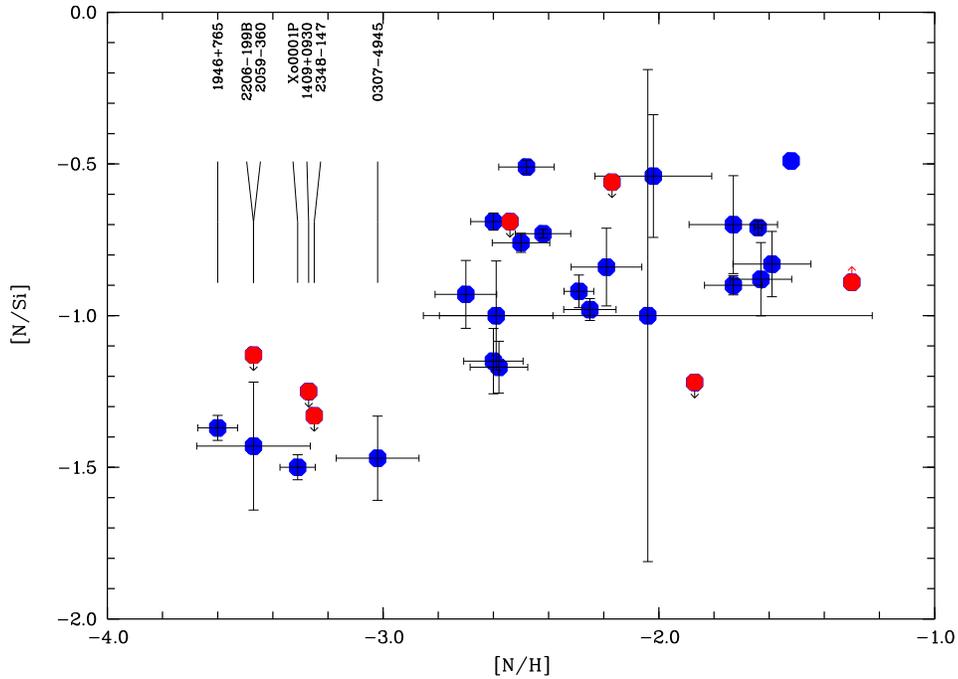}{6cm}{-90}{50}{50}{-200}{300}
\caption{[N/Si versus [N/H]} 
\end{figure}

  Fresh clues for the interpretation of the low N/O plateau can be obtained   plotting 
   the [N/Si]  values with respect to [N/H] as shown in Fig 4.  The distribution does not 
  show   a degeneracy with the metallicity any longer but reveal
   two different regimes with a transition occurring at    [N/H]$\approx$ $-$3. The    low N/O are in the region with [N/H]
$<$ -3  and the  high N/O for  [N/H]
$>$ -3.  Since there
   is no  obvious link between N  yields  and the  amount of   N content in the systems 
     we  think that what we see is most likely a  temporal  effect. In fact, since the O 
     abundance is more sensitive to  the SFR  history 
     while N depends from IMS evolutionary time-scales, 
     N could be a better indicator of the age of system.  
 Thus the two  regimes   reflect  probably  two  epochs. 
 
 We propose that the  higher plateau is produced by IMS
after  they had time to evolve  and  release    their N, while the lower plateau is produced by 
 more massive, namely M$>$ 8 M$_{\sun}$,   and relatively younger stars.   
At later times (or [N/H]$\approx$ -3) the contribution from the massive stars is 
overwhelmed by that from the IMS. In fact a hint of connection between the two plateaux can be recognized  in the  
 two DLAs which lie just below the primary plateau
at [N/Si] $\approx$ -1.1 and [N/H]$\approx$ -2.8 as shown in Fig 4. No objects are expected to be  found below the low N/O
 plateau and therefore it would be  rather crucial to improve the three upper limits 
to the levels at which they can probe the existence of the plateau. We note that for the DLA at z$_{abs}$=2.456 towards QSO 1409+095
  the upper limit is  only 0.17 dex  above the plateau and  in fact there is a 
hint of the presence of the NI 1200 \AA lines (Pettini et al 2002, their Fig 2).    

Standard models  do not predict N synthesis by massive stars but the situation is far from  being settled (Heger and  Woosley 2002).
 Chieffi et al (2001)    
 found that also Pop III stars of  intermediate mass  do experience thermal pulses and the third dredge-up  arguing that
stars with primordial composition are  sources of  C and N.
If we take the models of Limongi and Chieffi (2002) for zero metallicity, computed with FRANEC code CF85, 
 with  no mix of the H shell  and a mass loss parameter of $\eta$=6,
and we integrate with a Salpeter IMF  over the whole mass spectrum 4-80 M$_{\sun}$ we get a [N/O]=-0.85, namely very close to the higher plateau.
On the other hand the integration  in the mass range between 15 and 80 M$_{\sun}$ gives  [N/O]=-1.5, 
  which is very  close to the lower  plateau.
The models for low metallicity stars with rotational mixing or overshooting of Meynet and Maeder (2002)  predict 
 N production  for massive stars and  can also reproduce the 
  ratio observed in the lower plateau.  
 The   observations of DLAs might actually offer 
crucial  guidelines  for 
N production in massive stars.

The   fraction of systems on the two plateaux and their relative abundances  can be envisaged as two important tests 
to probe the origin of the 
two plateaux.
Presenly the fraction  of the systems forming the low N/O plateau  is   of    25\% of the total.  We can compute the expected ratio
assuming that we need  roughly 250 Myrs for significant  N production and  
computing the  age of the systems from the formation epoch to the observed redshift by means of 
a  look- back time  $t$= 0.538[(1+z)/10]$^{-3/2}$ Gyrs. The formation epoch is unknown but 
 to get the observed statistics we need to place  it   at z$\approx$ 4, which is quite conceivable.
Few  systems with low N/O have  z$_{abs}$ of about 2-3  suggesting
a rather continuous formation of the DLA.
The case  of QSO 1202-0725 with high N/O and  z$_{abs}$=4.38  is rather remarkable and  pushes back 
the   epoch of star formation considerably. 
  If the synthesis of   N observed by  IMS require about 0.25-0.5 Gyrs then the time of  star formation is shifted 
  to  a redshift greater than  six.  
  
An   $\alpha$- element enhancement over iron-peak element is expected  in the case that massive stars 
are responsible for  the  low N/O plateau. 
Fe is  produced considerably by Type Ia SNae and should   lag O   similarly to   N.  
 In Fig. 5  the [Si/Zn] and [Si/Fe] are plotted versus [N/Si]  
 Unfortunately there are no [Si/Zn] values for   DLAs in the low N/O plateau and the    [Si/Fe] are between 0.2 and 0.4 
 which  is rather common among the DLAs.  
  In  absence of dust these ratios would imply  an $\alpha$ enhancement, but unfortunately  
     there is no Zn information, and the test is not presently conclusive.

\begin{figure} 
\plotfiddle{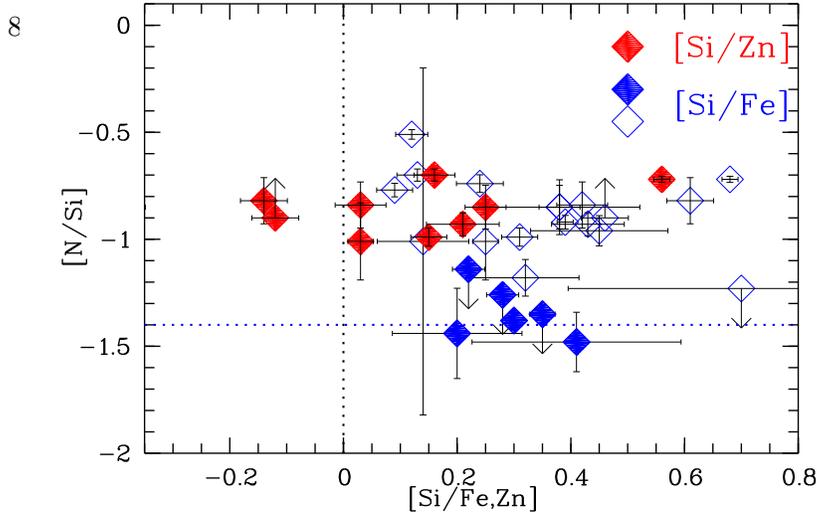 }{5cm}{-90}{55}{55}{-200}{300}
\caption{[N/Si] versus [Si/Zn] or [Si/Fe]. In the lower plateau there are only [Si/Fe]} 
\end{figure}

\section{ Acknowledgments} 
 
 It is a pleasure to acknowledge my collaborators, and in particular Miriam Centuri\'{o}n
  for allowing me to present here preliminary 
 results of common work.   Marco Limongi and Alessandro Chieffi are acknowledged  
 for making the computation  of N/O ratios with their model.

\end{document}